\documentclass[sigconf]{acmart}
\renewcommand\footnotetextcopyrightpermission[1]{}

\usepackage{amsthm}
\theoremstyle{definition}

\usepackage{amsmath}
\usepackage{subcaption}
\usepackage{graphicx}
\usepackage{pgfplotstable}
\usepackage{algorithm}
\usepackage{booktabs,siunitx}
\usepackage{algpseudocode}
\usepackage{threeparttable, tablefootnote}
\usepackage[normalem]{ulem}
\usepackage{tikz}
\usepackage{pifont}
\usepackage{multirow}
\usepackage{adjustbox}
\usepackage{multirow}
\usepackage{listings}

\newcommand{\HIDE}[1]{}

\newcommand{\ivxr}{iv4xr}
\newcommand{\IVXR}{Iv4xr}
\newcommand{\lr}{Lab Recruits}

\AtBeginDocument{%
  }

\acmConference[ATEST]{13th Workshop on Automating TEST case Design, Selection and Evaluation}{November,  2022}{Singapore}

\begin{document}

\title{An Agent-based Approach to Automated Game Testing: an Experience Report}

\author{I. S. W. B. Prasetya}
\orcid{0000-0002-3421-4635}
\affiliation{%
  \institution{Utrecht University}
  \country{the Netherlands}
  }
\email{S.W.B.Prasetya@uu.nl}


\author{Fernando Pastor Ricós}
\orcid{0000-0002-5790-193X}
\affiliation{
  \institution{Universitat Politècnica de València}
  \country{Spain}
  }
\email{fpastor@pros.upv.es}


\author{Fitsum Kifetew}
\orcid{0000-0003-1860-8666}
%
\author{Davide Prandi}
\orcid{0000-0001-9885-6074}
\affiliation{%
  \institution{Fondazione Bruno Kessler}
  \country{Italy}
  }
\email{{kifetew, prandi}@fbk.eu}

\author{Samira Shirzadeh-hajimahmood}
\orcid{0000-0002-5148-3685}
\affiliation{%
  \institution{Utrecht University}
  \country{the Netherlands}
  }
\email{S.shirzadehhajimahmood@uu.nl}

\author{Tanja E. J. Vos}
\orcid{0000-0002-6003-9113}
\affiliation{%
  \institution{Open Universiteit and Universitat Politècnica de València}
  \country{The Netherlands and Spain}
  }
\email{tanja.vos@ou.nl, tvos@vrain.upv.es}

\author{Premysl Paska}

\author{Karel Hovorska}
\affiliation{%
  \institution{GoodAI}
  \country{Czechia}
  }
\email{karel.hovorka@goodai.com}



\author{Raihana Ferdous}
\author{Angelo Susi}
\orcid{0000-0002-5026-7462}
\affiliation{%
  \institution{Fondazione Bruno Kessler}
  \country{Italy}
}
\email{{rferdous,susi}@fbk.eu}

\author{Joseph Davidson}
\affiliation{%
  \institution{GoodAI}
  \country{Czechia}
  }
\email{joseph.davidson@goodai.com}


\renewcommand{\shortauthors}{Prasteya et al.}

\begin{abstract}
   Computer games are  very challenging to handle for traditional automated
   testing algorithms. In this paper we will look at intelligent agents as a solution. Agents are suitable for testing games, since they are reactive
   and able to reason about their environment to decide the action they want to take.
   This paper presents the experience of using an agent-based automated testing framework called \ivxr\ to test computer games. Three games will be discussed, including a sophisticated 3D game called Space Engineers. We will show how the framework can be used in different ways, either directly to drive a test agent, or as an intelligent functionality that can be driven by a traditional automated testing algorithm such as a random algorithm or a model based testing algorithm.
\end{abstract}


\maketitle

This is a preprint of a paper with the same title. It is published in
the 13th Workshop on Automating TEST case Design, Selection and Evaluation (ATEST),
2022. The finalprint is published by ACM and can be found here:
\url{https://doi.org/10.1145/3548659.3561305}

\section{Introduction}

Computer games are notoriously hard to test automatically. Imagine we want to test some specific state in a computer game. To do this, the tester may need to guide an in-game character through thousands of fine grained interactions to arrive in the state of interest; only then the tester can check one or more assertions on that state. 
In other types of interactive systems, such as web or mobile applications, testers can use a record and replay technology to automate the execution of tests. A tester would record manual sessions where he/she interacts with the application under test. The recorded sequence of interactions are then used as test cases by replaying them. Unfortunately this works poorly, and even more so in the game setup as recorded game plays are very fragile. E.g. non-determinism would break recorded game plays, which is problematical because this is prevalent in computer games, e.g. due to their randomized logic or presence of concurrent components. Furthermore, if the game designer changes the layout of the game world, or the placement of some game items just a little, which happens frequently during the development, these also break recorded tests. 
So, to robustly automate the execution of test cases, we need a solution that possess reactivity (ability to react to unexpected changes) and some "intelligence" to autonomously re-plan the test sequence if needed, e.g. if the world layout has changed. There is one programming paradigm that allows these to be programmed naturally, namely agent programming. As such this makes an agent programming framework a good candidate to be used as the base to build an automated game testing solution -- \ivxr \ is such a framework.

The \ivxr\ framework provides a Java implementation of test agents \cite{prasetya2020aplib}.  A test agent can be connected to a game under test through an interface, and then used to autonomously drive a player character to do automated play testing.  The framework provides concepts such as 'tactic' and 'goal structure' to abstractly program reactive behavior and complex testing tasks. The framework also provides automated path finding and terrain exploration 
\cite{prasetya2020navigation} to enable a test agent to autonomously find a target game object it wants to test, regardless the layout of the game world.

This paper presents our experience of using \ivxr\ framework to test three different games: a Nethack-like 2D game called MiniDungeon, a 3D game called Lab Recruits, and a commercial 3D game called Space Engineers. Each will show a different use case of \ivxr. 

This paper is structured as follows.
Section \ref{sec.iv4xr} gives a brief overview of \ivxr\ and its agent programming.
Section \ref{sec.cases.overview} gives an overview of the three case studies that we will present; Sections 
\ref{sec.minidungeon},
\ref{sec.SE}, and
\ref{sec.LR} discuss each in more details.
Section \ref{sec.relatedwork} gives a brief overview of related work, and finally Section \ref{sec.concl} concludes.

\section{The \ivxr{} Test-agent Framework}\label{sec.iv4xr}

\begin{figure}[t]
\includegraphics[scale=0.27]{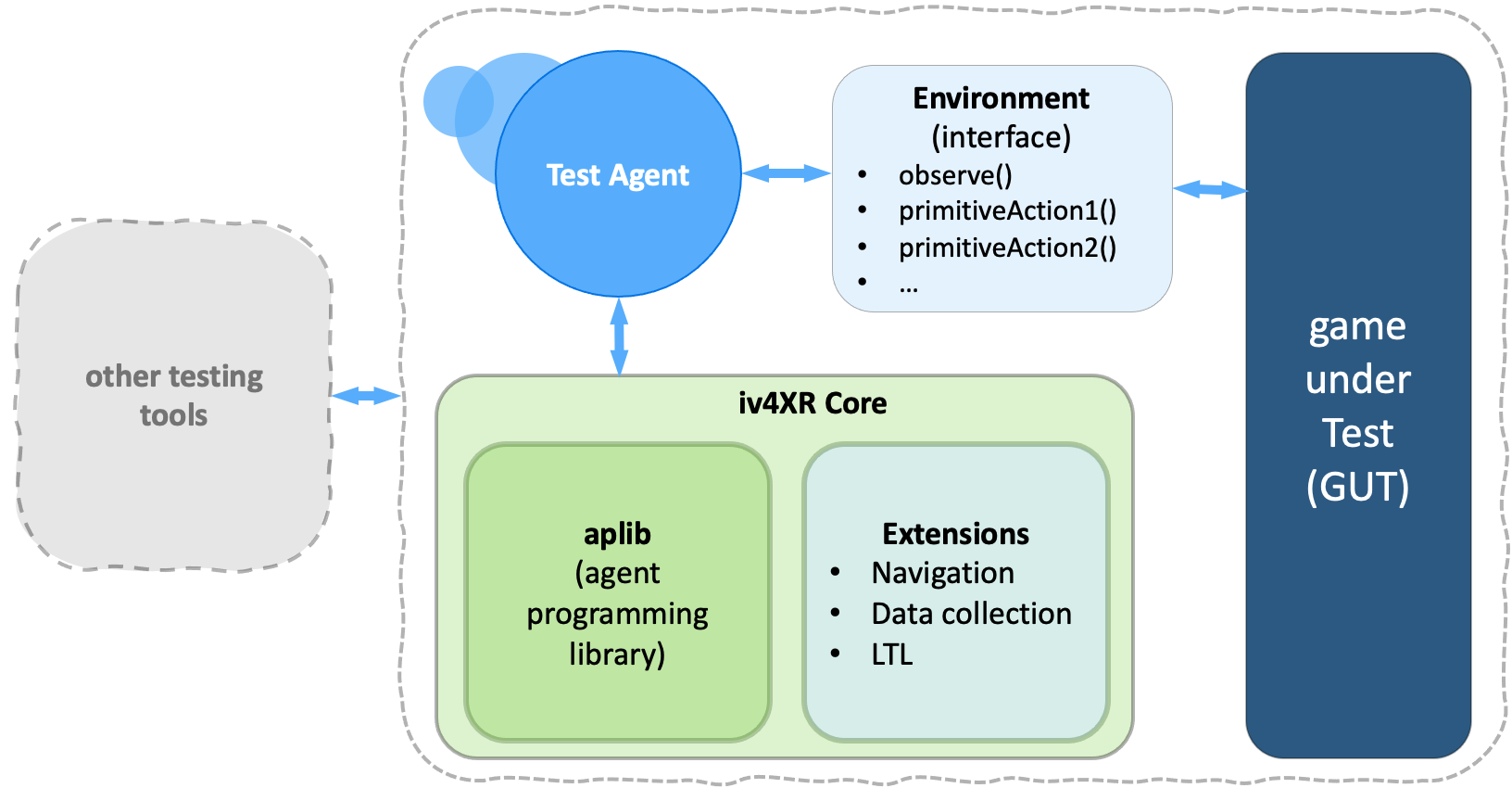}
\caption{A high-level architecture of \ivxr\ framework.}\label{fig:arch1}   
\end{figure}


The high-level architecture of the \ivxr\ framework is shown in Figure \ref{fig:arch1}. 
A test agent can be connected to a game under test (GUT) through an interface called {\em Environment}. The game developers should implement this interface. It should provide a method ${\sf observe()}$ to observe the state of the game, and methods implementing primitive actions the agent can do on the game, such as interacting with a nearby game object, or moving to a certain direction for some unit of distance. What actions are to be provided, and how much observation ${\sf observe()}$ reveals, are up to the developers to decide. However, there are some aspects to consider. For example, allowing the agent to instantly access to the state of all game objects would make testing easier, but this might be computationally excessive as the GUT might then need to send over the states of thousands of objects to the agent. We also lose some realism as actual players can only see objects visible in their screens. An all-seeing agent might take actions that actual users would not do.

In the most basic form, a test agent drives the GUT by invoking its Environment's actions and samples the GUT's state as it goes to check if the GUT is in the correct state.  \IVXR\ is inspired by so-called BDI (Belief-Desire-Intent) agents \cite{dastani20082apl,bordini2007programming,rao1991modeling}. This type of agents has a quite different execution model than a traditional procedure, so it is useful to first explain this. A BDI agent has a set of goals and runs in so-called {\em deliberation} cycles
until its goal set is empty. At every cycle, the agent observes its environment, decides which action to do, and executes the action.
It also decides if the current goal is accomplished, or if it should be dropped, and if so, which goal to pursue next.
Specific for an \IVXR\ agent, it accumulates all observations it gets so far into what can be seen as 'belief': the latest observation is factual, but older observations in the belief may no longer be valid in the actual GUT state. 
A BDI agent is allowed to act on belief, e.g. if an object $o$ exists in its belief,
it can optimistically decide to go to $o$, believing it still exists in the actual game world.
The fact that a BDI agent runs in deliberation cycles also makes it highly {\em reactive}, as it allows the agent to continuously, or at least  frequently, sample the state of the GUT and immediately acts after each sampling, which makes it very suitable for controlling a game. 

The basic form of BDI agent programming is to specify which action to select at each deliberation cycle, which can be expressed {\em declaratively} with guarded actions a la Action System \cite{UNITYcm}. 
The snippet below shows an example of how this looks like in \ivxr\ (some concrete syntax are omitted). $B$ represents the agent's belief; $B \rightarrow expr$ is a lambda expression that, here, represents an action.
\[
{\bf var}\ \begin{array}[t]{l}
  tactic_1 \ = \ {\sf ANYof}( \\
   {\sf action}().{\sf do1}(B \rightarrow B.env().moveUp()).{\sf on}(g_1)\; , \\
   {\sf action}().{\sf do1}(B \rightarrow B.env().moveDown()).{\sf on}(g_2)\; , \\
   ... \\
   {\sf action}().{\sf do1}(B \rightarrow B.env().useHealKit()).{\sf on}(g_k)\; 
   )  
  \end{array} 
\]
where $moveUp()$, $moveDown()$, $useHealKit()$, etc are methods we can imagine as provided by the Environment $env()$, and $g_1 .. g_k$ are guards specifying when the corresponding action is enabled for execution (e.g. $g_1$ could require that the way upwards is clear). Only enabled actions are executable; if there are more than one, the $\sf ANYof$ combinator will select one randomly. 
In \ivxr, a system of actions such as the one above is called a {\em tactic}.  Given a tactic, an agent will keep executing it until its current goal is achieved (or it runs out of budget). E.g. this goal could be 'to obtain a key' (e.g. because we want to check its properties).

If we remove all the guards, the tactic above would be how we can program a random test agent. 
Guards add some intelligence in choosing better actions (than just randomly), e.g. the action $useHealKit()$ can be guarded so that it becomes enabled when the character health drops under a certain critical level. 

\subsection{Navigation} \label{sec.navigation}

\begin{figure}[h]
\includegraphics[scale=0.22]{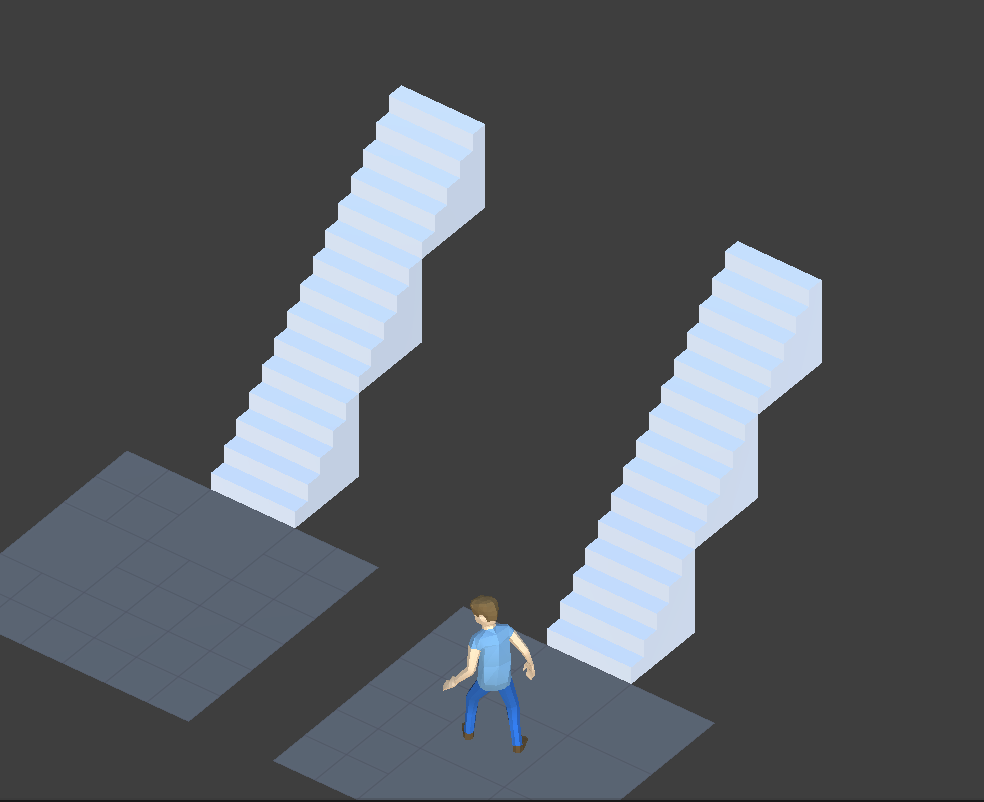}
\includegraphics[scale=0.22]{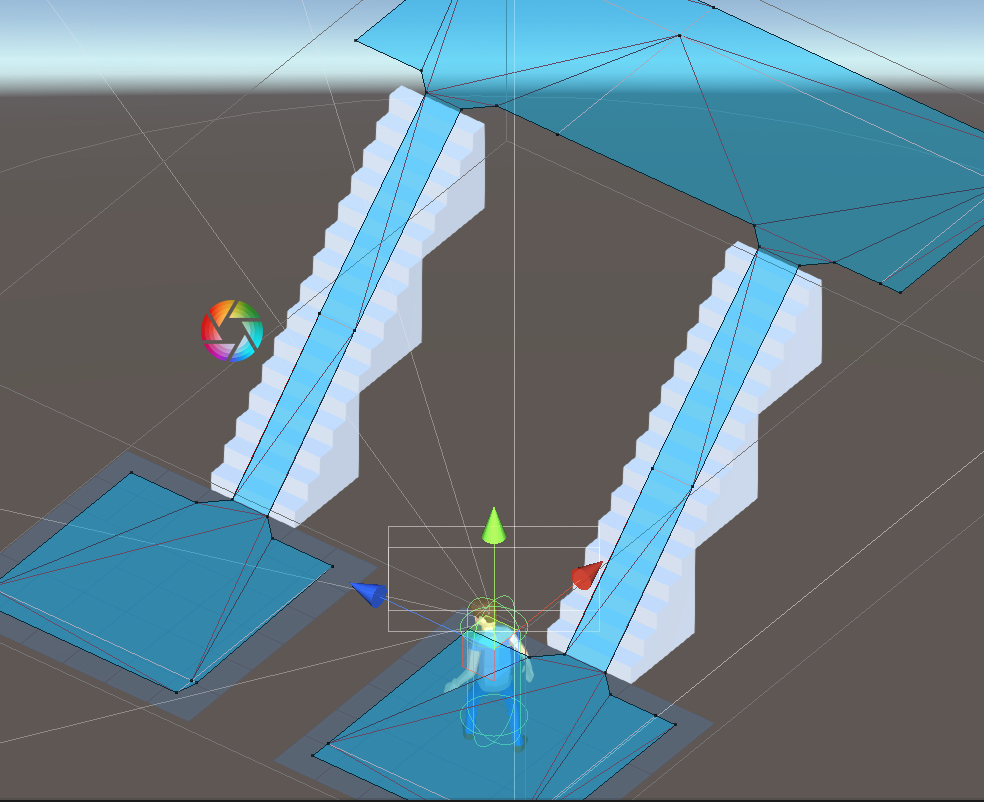}
\caption{\it The picture to the right shows the mesh (blue surface), consisting of triangles (edges colored red), in a UNITY game.}\label{fig:mesh}   
\end{figure}

A basic, but important, task that should be automated is navigation. 
The previous $tactic_1$ can do it, but not effectively. 
In fact, navigation in a game wold is usually non-trivial due to its complex layout 
and presence of dynamic obstacles. 
A standard solution is to represent walkable parts of the game world, which can be an infinite continuous space, as a finite {\em navigation graph}, after which a path finding algorithm such as A* \cite{millington2019AI} can be applied to guide the agent to get to a target location. \IVXR\ provides several ways to do this reduction. For example if the GUT can export a so-called {\em navigation mesh}, \ivxr\ can convert it to a navigation graph. Figure \ref{fig:mesh}  shows an example of such a mesh in a game engine called UNITY.
A mesh is a finite set of connected triangles that cover a walkable surface. As such, it induces a  navigation graph. If the GUT does not produce a navigation mesh, \ivxr\ can also construct a navigation graph on the fly, based on the geometry of the objects an agent sees. 

From this, two tactics can be constructed \cite{prasetya2020navigation}. First, $navigateTo(o)$, which when repeatedly executed would guide the agent to reach the location of an object $o$, if the location is known. Second, $explore()$ to guide the agent to the closest unexplored area of the game world. 
So, rather than the previous $tactic_1$ we can now have the following, if the goal is to obtain some object 'key' $k$:

\[
{\bf var}\ \begin{array}[t]{l}
  tactic_2 \ = \ {\sf FIRSTof}( \\
   {\sf action}().{\sf do1}(B \rightarrow B.env().useHealKit()).{\sf on}(g)\; , \\
   navigateTo(k)\; , \\
   explore()) 
  \end{array} 
\]
Above we use a priority-based selector $\sf FIRSTof$ rather than the previous random selector $\sf ANYof$. The 
sub-tactic $navigateTo(k)$
will take the agent to the key $k$, if its location is known. Else the tactic is not enabled; $\sf FIRSTof$ will instead choose $explore()$ to explore the world until the agent sees $k$. Though, if its health drops too low, it will first use a healing kit to fix itself. Note that the last  adds {\em reactivity} to handle an 'emerging situation', namely when the health drops too low. This can be extended to handle more emerging situations, such as approaching enemies,
and thus equipping the test agent with some logic to make it more adept in surviving the game  (useful, as a dead agent can't perform testing tasks).


\subsection{Formulating a Testing Task: Goal Structures}\label{sec.goalstructure}

An \IVXR\ agent can be given multiple goals. Unlike other agent programming languages, \ivxr\ requires the goals to be structured. A {\em goal structure} is tree with goals as leaves and control-combinators as nodes, specifying either an order or a priority with which its subgoals are to be solved. For example a sequential testing task to find the key $k$, to pick it up, and to check that it can be used on door $d$ can be formulated as a goal structure such as the one below:
\[ {\sf SEQ}(
   \begin{array}[t]{ll}
   \mbox{"$k$ is found"} & .{\sf withTactic}(T_1(k)), \\
   \mbox{"$k$ is picked up"} & .{\sf withTactic}(T_2(k)), \\
   \mbox{"$d$ is found"} & .{\sf withTactic}(T_1(d)), \\
   \mbox{"$k$ is used on $d$"} & .{\sf withTactic}(T_3(k,d))) \\
   \end{array}
\]
$\sf SEQ$ requires its subgoals to be solved in the order they are given.
For more combinators, including conditional and repetition, see \cite{prasetya2020aplib}, 
with which even a test {\em algorithm} can be expressed, e.g. when the exact sequence of sub-tasks is not known upfront \cite{shirzadehhajimahmood2021}.

\subsection{Integration with Other Testing Tools}

Another value of the \ivxr\ framework is to be used as a rich adapter to enable traditional automated testing tools to target computer games (see also the architecture in Fig. \ref{fig:arch1}).
For example we used this scheme to allow a GUI-testing tool TESTAR \cite{Vos+2021} and a model-based testing (MBT) tool to target games in two of our case studies.
E.g. TESTAR exploits \ivxr\ Environment and navigation graph to perform smart monkey testing.
The MBT tool can efficiently generate test cases from an EFSM model of a game, which subsequently are translated to goal structures for a test agent to execute. This is a very simple integration scheme. A translator must indeed be written, but this only need to be written once for each game.


\section{Cases Overview}\label{sec.cases.overview}

In the coming sections we will discuss our experience in using \ivxr\ for testing three different games: MiniDungeon, Space Engineers, and Lab Recruits. With each we also want to show a different way of using \ivxr. In the MiniDungeon case we will show a {\em direct use} of a \ivxr\ test agent  to  check a set of  correctness properties of the game, such as the reachability of key objects in the game. The game has much randomness in its logic and enemies too. The test agent needs to be able to deal with both to remain robust and survive long enough to complete its testing tasks.

In the Space Engineers case we will show a setup where \ivxr\ is leveraged to enable another testing tool, in this case the GUI testing tool TESTAR \cite{Vos+2021}, to do automated exploratory game testing.

In the Lab Recruits case shows a setup where \ivxr\ agents are used as an intelligent executor for model based testing (MBT). This setup allows the behavior of a game under test to be described abstractly using e.g. an extended finite state machine (EFSM), where the concrete layout of the game world can be abstracted away from the model. An MBT algorithm can very efficiently generate abstract test cases from such a model. These abstract cases are fed to the agents that will carry them out. The agents exploit automated navigation and exploration from \ivxr\ (Section \ref{sec.navigation}) to explore the concrete world layout to find the game objects in the test cases.

\section{MiniDungeon}\label{sec.minidungeon}

\begin{figure}[h]
\includegraphics[scale=0.3]{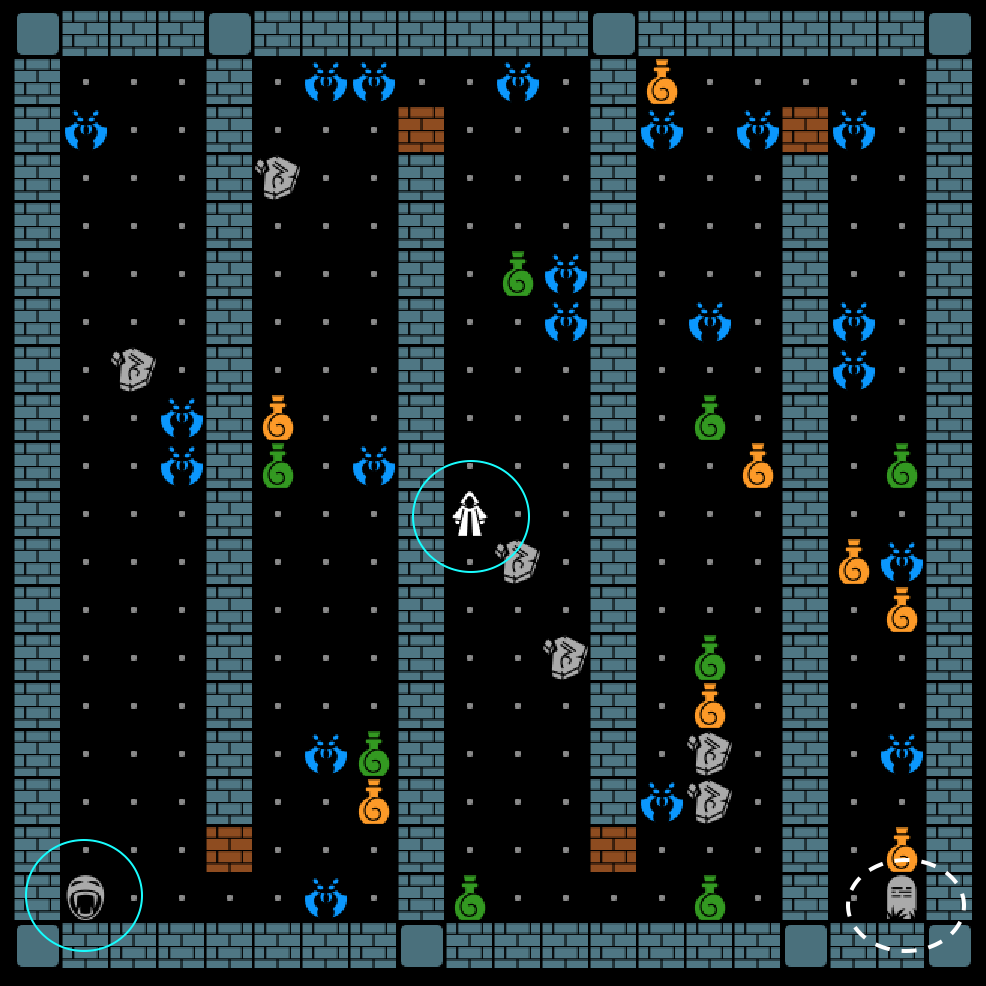}
\caption{The MiniDungeon game.}\label{fig:minidungeon}   
\end{figure}

MiniDungeon is a small 2D, turn-based, Nethack-like game written in Java. Figure \ref{fig:minidungeon} shows a screenshot.
The game can be played by a one or two players. Figure \ref{fig:minidungeon} shows two players (circled blue).
The players' goal is to cleanse the shrine (circled white). To do so, a scroll is required (gray icon) which a player must bring to the shrine. Only a holy scroll will cleanse the shrine, but the player does not know which scroll is holy until he/she tries it. If the shrine is cleansed, it becomes a portal that takes the player to the next level. This goes on until the final level; cleansing the shrine there wins the game.
Along the way there are monsters (blue figures) that can hurt players, but also potions to help them. A greedy strategy that just collects all scrolls and potions does not work because the player's bag has limited space, which is either 1 or 2.

\subsubsection*{The case}
The source code is about 1.2K lines. There are unit tests providing good coverage for their respective targets. However, in total they only cover 20\%  of the whole code base because a large part of the game is simply hard to unit test.

Monsters, shrines, scrolls, potions, and even players are so-called {\em game objects}. As common in game implementations, MiniDungeon has a so-called {\em game loop}, where at every iteration it executes the entire state update for each new turn; so, executing the players' move and all monsters' move for the turn. Game objects may have multiple properties, but they do not have much behavior on their own. For example, when a monster move, the logic that decides where to move resides in the game loop. This cannot be delegated to the monster-object itself, since the latter only knows its own state and does not know which neighbouring squares are empty to move to. So essentially, most of the game logic resides inside the game loop. Unfortunately this game loop is hard to unit-test.

For example, the game loop moves the monsters in random directions. 
Imagine we want to verify that it will never move a monster to an occupied square. To do this with the usual unit-testing setup we will need to create 
a set of test fixtures in the form of a game level, seeded with at least one monster and different game objects in different neighboring squares. This is combinatorial in nature, so it takes substantial effort to hand-craft the fixtures. Then, to test the monster's move, we run a the game loop for one ore more turns on all the fixtures, and repeated multiple times using different random seeds to account for the randomness in the move's logic.
A more practical approach is to implant the check as an assertion inside the game; a snippet of this is shown below:

\begin{lstlisting}[
  morekeywords={assert},
  xleftmargin=10pt,
  frame=l
  ]
// move the monster to sq:
...
if(Debug.ON) assert(world[sq.x][sq.y]==null) ;
world[sq.x][sq.y]=m;
\end{lstlisting}

\noindent
And then we simply 'play' the game several times. The implanted assertion will catch erroneous monster-moves. This does not require manually creating fixtures. We also want to do the play testing automatically. To achieve this we program an automated play testing agent using \ivxr. 

\subsection*{\IVXR\ solution}
We first implement the Environment component in Figure \ref{fig:arch1} that serves as the interface between the agent and the GUT. Its main APIs is shown below:

\begin{lstlisting}[
  xleftmargin=10pt,
  frame=l
  ]
class MyAgentEnv extends Iv4xrEnvironment{
   WorldModel observe(agentId)
   WorldModel command(agentId,cmd) 
}
\end{lstlisting}
The method ${\sf command}(a,cmd)$ simulates a key pressed by a player as a command. E.g. the key 'w' and 's' cause the agent $a$ to move up respectively down, whereas 'q' causes the game to end.

The method ${\sf observe}(a)$ returns what the agent $a$ currently sees. In Fig. \ref{fig:minidungeon} we artificially set the view distance to $\infty$, but normally the view is limited, e.g. only 3 squares away. 
Observation is represented as an \ivxr\  datastructure called $\sf WorldModel$. Essentially, it is a set of 'entities', each representing a game object as a record of a unique ID, timestamp, its physical location, and a list of name-value pairs describing the object's other properties/state. Entities may have sub-entities if needed. 
As mentioned, the agent automatically accumulates observations into its belief, also represented 
as a $\sf WorldModel$. In this belief, when e.g. a monster was observed, the observation is kept even if the monster is no longer visible. It is maintained until a new observation updates the monster's state, or if the new observation says the monster has been destroyed. This belief gives the agent more information/depth for making its decision, rather than just reacting to what it currently sees.

\subsubsection*{Automated exploration and navigation}

We cannot do much automated play testing if the agent is not able to autonomously explore a game level and navigate to game objects. To support we implement the tactics $explore()$ and $navigateTo()$ discussed in Section \ref{sec.navigation}. \IVXR\ provides the main worker functions that do the calculation over the navigation graph, but the calculation results still need to be translated to calls to actual movement actions as provided by GUT through the Environment.
This requires some programming work, but not much (40-50 lines).

\subsubsection*{Survivable play testing} When the GUT has hazards (such as aggressive monsters), just automated navigation as above is not sufficient. The agent must be smart enough to handle the hazards in order to live long enough to reach the state it is tasked to check. Programming a survivable play testing agent takes  more effort.

Imagine a goal $entityInCloseRange(o)$ that is achieved when the agent reaches a square next to the object $o$. To solve it, rather than simply using $navigateTo(o)$, we use a tactic similar to $tactic_2$ from Section \ref{sec.navigation}, extended with more combat sub-tactics, e.g. 
line \ref{line.attack} to attack a monster that engages the agent and line \ref{line.rage} to quaff a rage potion to increases the agent's power when in combat.

\begin{lstlisting}[
  morekeywords={assert},
  xleftmargin=10pt,
  frame=l,
  mathescape=true,
  numbers=left
  ]
FIRSTof(
  useHealingPot().on_(hasHealPot_and_HpLow), $\label{line.heal}$
  useRagePotAction().on_(hasRagePot_and_inCombat), $\label{line.rage}$
  attackMonster().on_(inCombat_and_hpNotCritical), $\label{line.attack}$
  navigateToTac(o),
  explore(),
  ABORT())
\end{lstlisting}

\subsubsection*{Programming a complete playtest}

As an example of a complete playtest we configure MiniDungeon to generate a game world consisting of $N{=}2$ levels. We program the test agent to first cleanse level-1's shrine, and then that of level-2, which would then wins the game for the agent. A playtest does not have to be winning though, but a winning playtest usually covers most key features of the game. To cleanse a shrine the agent will have to try different scrolls until it gets the right one. We do not want to explicitly program the sequence of scrolls to try. This would make the test less robust. Instead we use an implementation of the online search algorithm in  \cite{shirzadehhajimahmood2021} to let the agent autonomously do the search.
The algorithm will require some components to be 'plugged-in', such as the goal $entityInCloseRange(o)$ mentioned before, but also a goal to make $o$ interacted (after the agent stands next to it).
Once the algorithm is set up, we can use it to automate a task of the form:
\[ solver(a,T,o,\phi) \]
This constructs a goal structure for an agent $a$, that seeks to change the state of object $o$ to a new state satisfying $\phi$. In our case, $o$ is a shrine and $\phi$ is "$o$ is cleansed". It does this by autonomously searching objects of type $T$ and then using them, one at a time, until the aforementioned goal is accomplished.

So now the whole playtest can be written, essentially, just as:
\[ {\sf SEQ}(
   \begin{array}[t]{l}
   solver(a,Scroll,shrine_1, "shrine_1 \mbox{\ is cleansed}"), \\
   interacted(shrine_1), \\
   solver(a,Scroll,shrine_2, "shrine_2 \mbox{\ is cleansed}")) \\
   \end{array}
\] 
The test passes if this goal structure is solved. Additionally, various assertions over the agent's belief are also checked, e.g. that the agent health is expected to eventually drop below its maximum (as there are monsters attacking the player) but it never drops to 0, that the number of items in its bag never exceeds the bag's capacity, that the agent never walks through a wall, and so on.

\subsection*{Experience}

Table \ref{tab.result.minidungeon} shows the improvement we get. Without agent, the unit tests only covers 18.8\% ($U_{cov}$) of the total code base . With the agent play tests we can cover 88.4\% ($all_{cov}$), which is a huge improvement. Furthermore the latter found two bugs that were not found by unit testing. These are subtle bugs that are difficult to catch at the unit level unless we specifically were looking for them. For example one of the bugs occurs when the two players are in different levels but the $xy$-projection of their visibility areas overlap. The bug caused some squares that should be visible to a player to be missed. Such a situation is just hard to anticipate at the unit level.

\begin{table}
{\small
\begin{tabular}{lccc} \hline
  & 
  \multicolumn{2}{c}{$locs$} 
  & $cc$ 
  \\ \hline
MiniDungeon (GUT) & & 1196 & 325 \\
iv4xr-lib (playtest infrastructure)  & & 869  & 219\\
\ \ \ Environment implementation & 
   177 \\
\ \ \ Tactic and goal lib & 426 \\
\ \ \ Utils and other & 266 \\
unit tests      
  & & 236 & 129\\
agent playtests 
  & & 196 & 74 \\ \hline
\\
\end{tabular}
}
\caption{\it The GUT, tests and test-infrastructure size and complexity, given in, respectively, lines of code ($locs$) and cyclomatic number ($cc$). }\label{tab.effort.minidungeon}
\end{table}

\begin{table}
{\footnotesize
\begin{tabular}{lcccccccc}\hline
      & $C$ & $i$ & $cc$ 
      & $U_{cov}$ 
      & $PT_{cov}$
      & $all_{cov}$
      & $U_{bug}$
      & $PT_{bug} $\\ \hline
Entity & 
   11 & 360 & 21 
   & 81\% & 97.8\% & 100\% 
   & 2 \\
Maze &
   1 & 320 & 18 
   & 89.9\% & 95\% & 95\% 
   & 1 \\
MiniDungeon &
   2 & 2282 & 197 
   & 14.2\% & 84.4\%  & 84.5\% 
   & & 2 \\
MDApp &
   1 & 1228 & 89 
   & 0\% & 92.3\% & 92.3\% \\
All &
  15 & 4790 & 325
  & 18.8\% & 88.2\% & 88.4\% 
  & 3 & 2 \\ \hline
\\
\end{tabular}
}
\caption{\it
  The table shows some statistics of four classes that made the game MiniDungeon, and
  how well the tests cover them.
  $C:$ the number of classes that a top-level class has;
  $i:$ the number of instructions;
  $cc:$ cyclomatic complexity;
  $U_{cov}, PT_{cov}$: instruction coverage of respectively unit tests and play tests with an agent;
  $all_{cov}:$ the coverage of combined tests;
  $U_{bug}, PT_{bug}:$ the number of bugs found by respectively unit tests 
  and play testing with an agent.
}\label{tab.result.minidungeon}
\end{table}

Table \ref{tab.effort.minidungeon} gives some indication on the investment and effort needed to do agent playtesting. The library that provides a implementation of $Environment$ along with smart tactics and goal-structures is about 850 lines large, which substantial relative to the size of the GUT. Fortunately, it is less complex (see the $cc$ number). This part is a one-off investment. The playtests themselves are smaller and less complex than all the unit tests together.
Given the huge gain in the test coverage, and the convenience with which we can subsequently write automated playtests, the extra investment in the infrastructure is arguably well spent.


\section{Space Engineers: TESTAR-\ivxr} \label{sec.SE}

\begin{figure}[h]
\includegraphics[scale=0.3]{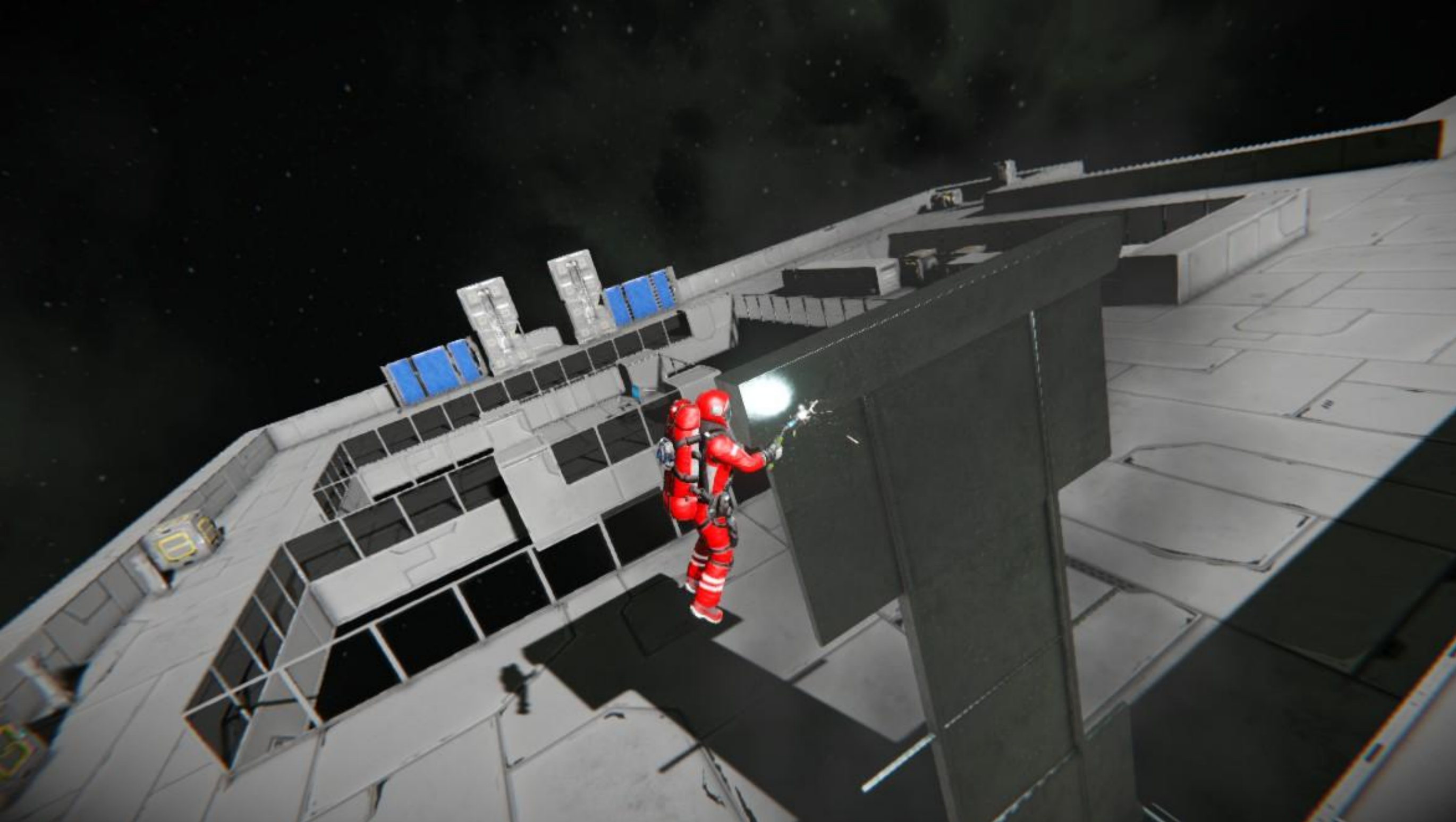}
\caption{\it 
An example of a Space Engineers level. 
}
\label{fig:SE}   
\end{figure}

Space Engineers (SE) is a complex open-world game developed by Keen Software House and GoodAI (GA). 
It offers users the simulation of a realistic 3D environment with volumetric physics and objects with mass, inertia, and velocity. 
SE users can use multiple types of blocks and tools to build any structure like bases or spaceships. 
Figure \ref{fig:SE} shows a construction in the space of the game. 
The testing process of SE consists of a team of testers who manually test the functionality and visuals of aspects of the game. 
Given the game's complexity, more than 10,000 manual tests are performed for each major release. 

TESTAR is an open-source tool for scriptless GUI testing that automatically generates test sequences of (state, action)-pairs at run-time. 
This scriptless tool does not follow previously created or recorded scripts or models to select which action to execute, but instead follows an action selection mechanism (ASM) to make decisions on the fly. 
The main advantage of this is that it is really a push and go approach to complement manual testing. 

A computer game like SE is however not a GUI application in the traditional sense, as it does not expose a widget tree that a typical GUI testing tool can target. 
To be able to target SE we exploit \ivxr{}, in particular its Environment and Core components. Similar to the MiniDungeon case (Section \ref{sec.minidungeon}), the Environment's $\mathsf{observe}()$ constructs a $\mathsf{WorldModel}$ that keeps track of active game objects. This $\mathsf{WorldModel}$ has the same role as a widget tree in a GUI and allows TESTAR to target the game objects it tracks.
We integrated TESTAR with \ivxr{} and use it as an exploratory test agent on SE \cite{pastor2022scriptless}. 
The logical flow of TESTAR consists of: connecting with the SE system, realizing an observation to obtain information about all existing virtual entities in a specific range, deriving the possible actions to execute, and selecting one to transit to a new state. 
There are two types of actions: basic commands and compound tactics. 
A basic command action is the most basic event we can execute in SE, e.g., move or rotate one step, equip a tool and start or stop using a tool. 
A compound tactic action contains several basic commands that simulate user decisions. 
For example, to interact with a block, we need to rotate the agent to aim at the block, move to reach the block, equip a tool and then start using this tool. 

\begin{figure}[h]
\includegraphics[scale=0.5]{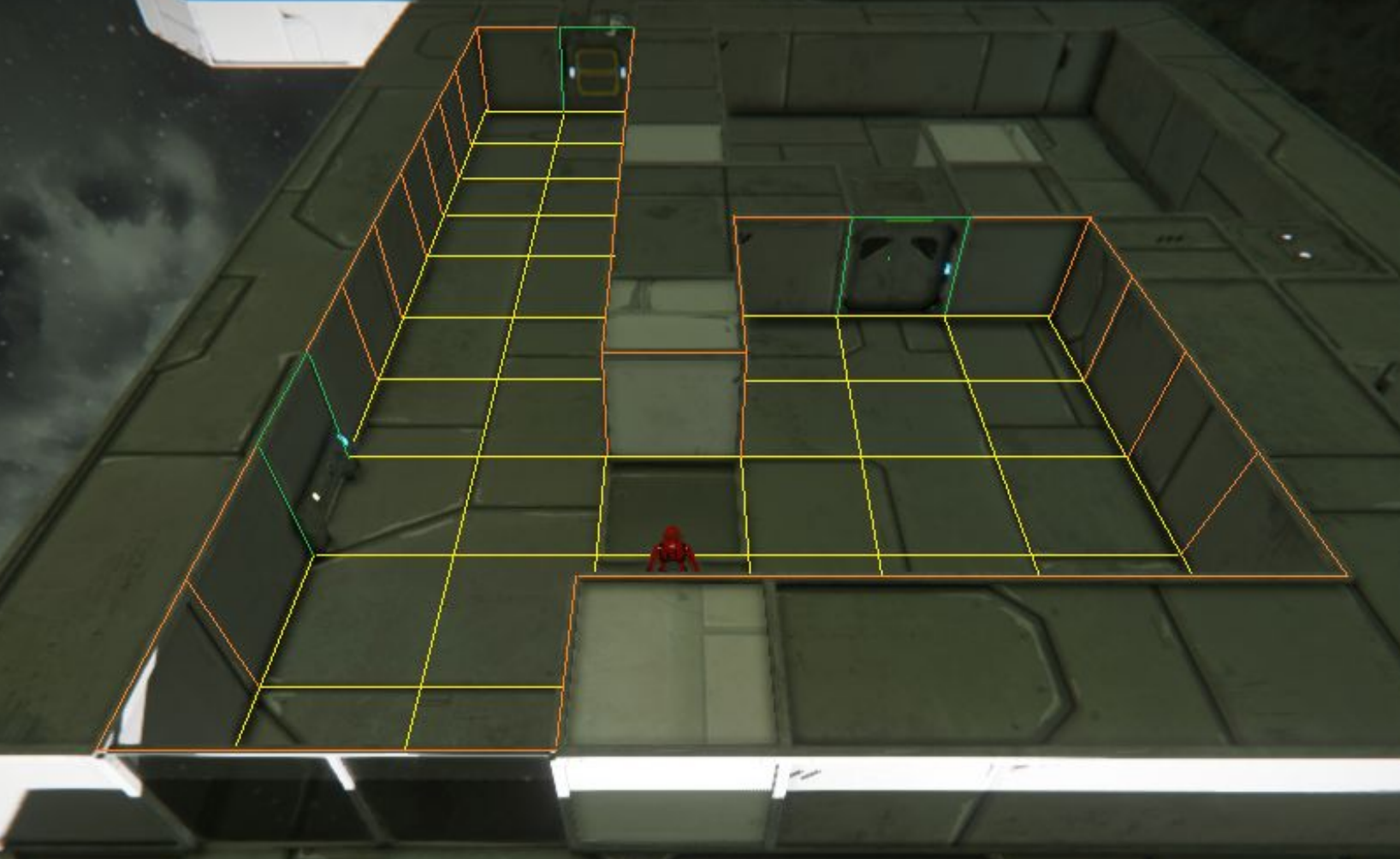}
\caption{\it 
An example of the agent observation and use of block geometry to compute navigable space in Space Engineers. 
The floor contains 2D navigable nodes (yellow lines). 
The walls require the use of a jet-pack to fly over them (orange lines). 
The agent can observe a series of interesting blocks to interact (green lines). 
}
\label{fig:SE_nav}   
\end{figure}

One of the main challenges in SE was to allow agents to calculate the navigation to a position to reach blocks. 
Because, unlike other game systems, SE does not have the functionality to produce a default navigation mesh. 
To deal with this, the \ivxr{} framework allows using the geometry information of the observed entities to construct a navigation graph on the fly, and calculate if the agent can follow a path of nodes to reach a position. 
TESTAR agent uses this feature to derive compound tactical actions, containing navigation, over the 3D space adjacent to an observed block. 
Figure \ref{fig:SE_nav} shows a representation of how the geometry of an SE level is used to calculate the navigable space. 

\subsection*{Experience}

A computer game like SE is very challenging for any automated testing tool to target, as the tool will also have to deal complex calculation to control terrain navigation and body motion, which most GUI testing tools are not equipped with.
Using the above setup with \ivxr, TESTAR can now do this. 
We then use the setup to do automated exploratory testing on SE. As the ASM policy we simply use guided random, where priority is given to interacting with game objects that have not been tried before.
The objective of the exploratory agent is to navigate to the interesting blocks, interact with them, and automatically validate their physics such as material integrity. 
%
We use TESTAR's feature for specifying oracles to strengthen the test by 
adding generic oracles to test the robustness of SE systems, e.g., to detect if the process crashes or hangs or to find exception messages in the application log.
Additionally, to test part of the game's functional aspects, custom oracles are also added, such as 
validating that the jet-pack settings are correct after interacting with functional blocks that move the agent. 

We apply this exploratory test on various levels of SE to verify the TESTAR agent functionality regarding navigable actions and oracles. 
A video of one of these tests can be seen here\footnote{\url{https://www.youtube.com/watch?v=ho1EMVtr8C4}} where TESTAR explored a small level for about three minutes and found a jet-pack bug. 

\HIDE{
\subsection*{Future work}
The integration of the \ivxr{} plugin allows TESTAR to act as an exploratory agent on the SE system. 
Currently, the Space Engineers and the iv4xr-TESTAR partners are collaborating to integrate new oracles that allow the detection of functional errors in the game and preparing an empirical experiment to validate the effectiveness of the scriptless exploratory agent. 
This validation will consist of using metrics to measure the code coverage and spatial coverage of the exploratory process, as well as preparing scenarios of previous bugged versions to validate the detection of errors. }

\section{Lab Recruits: MBT-\ivxr} \label{sec.LR}


In this example,  we exploit \ivxr\ to enable model based testing (MBT) of computer games.  MBT is widely used in industrial engineering, but applications in gaming are quite limited  \cite{iftikhar2015automated, ferdous2021search}. A major burden is the complex layout of the game world, which is very difficult to describe in the usual behavioral models (e.g. FSM) used in MBT. The \ivxr\ framework adds an intermediate abstraction level that provides navigation primitives and goal structures, allowing MBT to focus on the behavioural aspects of the game.


\begin{figure}[h]
\includegraphics[scale=0.2]{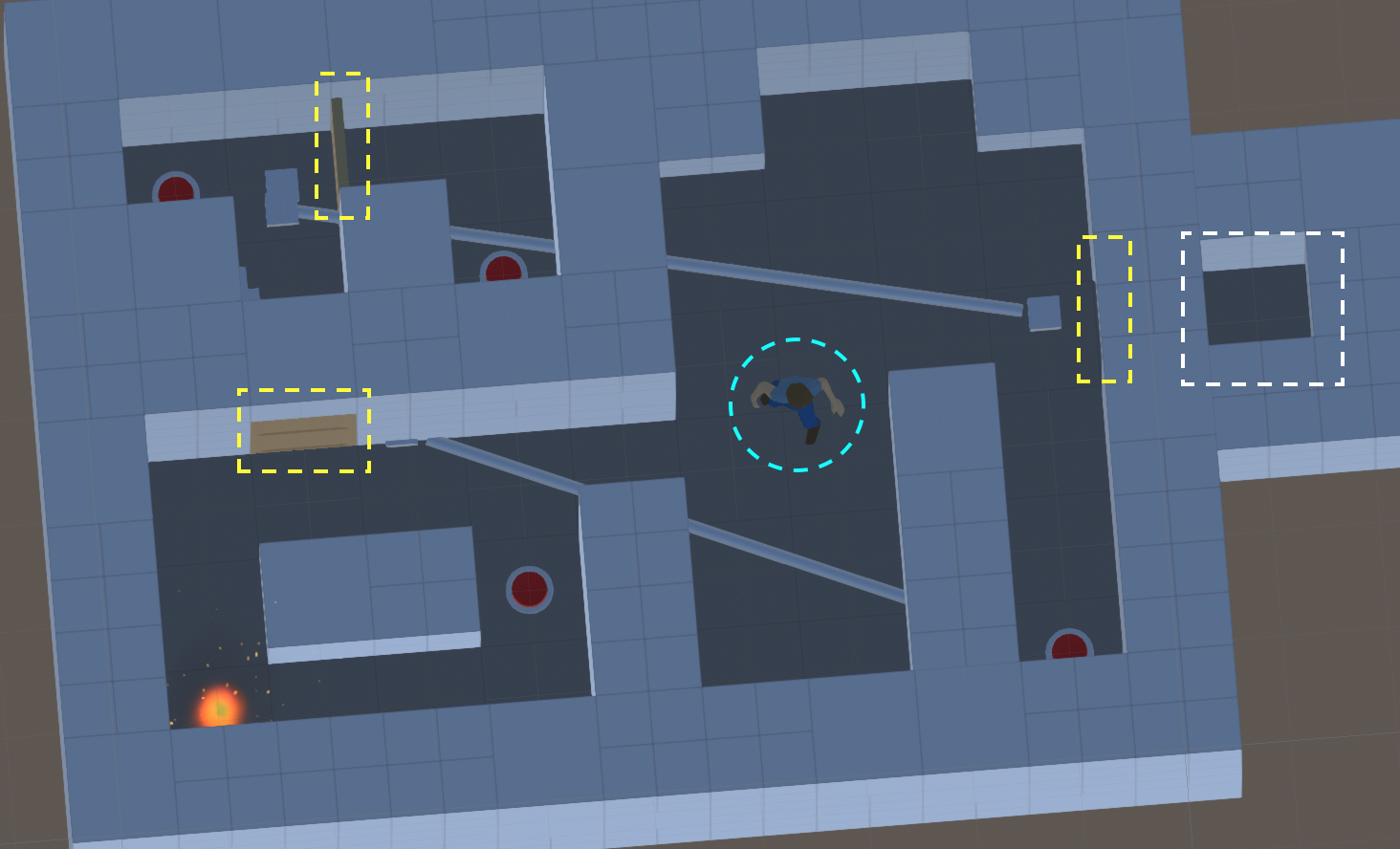}
\caption{\it A small level in the Lab Recruits game.
It has three rooms guarded by doors (marked yellow).
There are four buttons (red), each can toggle the state of zero or more doors. The player is marked by the blue circle. The goal of this level is to reach the white marked room.}\label{fig:LR}   
\end{figure}
Here, we apply MBT through \ivxr\ on \lr\footnote{\url{https://github.com/iv4xr-project/labrecruits}}, a 3D maze game.  The game allows players to explore a level, that typically consists of rooms guarded by doors, which can be opened by toggling the right buttons. The button-door connections are many-to-many, defining non-trivial paths that a player must discover to arrive at a given room. Other game objects include fire hazards and goal flags that give points and heal the player. A small \lr\ level is shown in Fig. \ref{fig:LR}. 
Game levels are defined as CSV files, which include the layout of the world and the (initial) placements of game objects. We exploit this to write a level generator~\cite{ferdous2021search}, capable of creating very large levels (which otherwise would be very labour intensive to craft by hand), along with an  Extended Finite State Machine (EFSM)~\cite{cheng1993automatic} that models the logic of the levels.

\begin{figure}[h]
\includegraphics[scale=0.25]{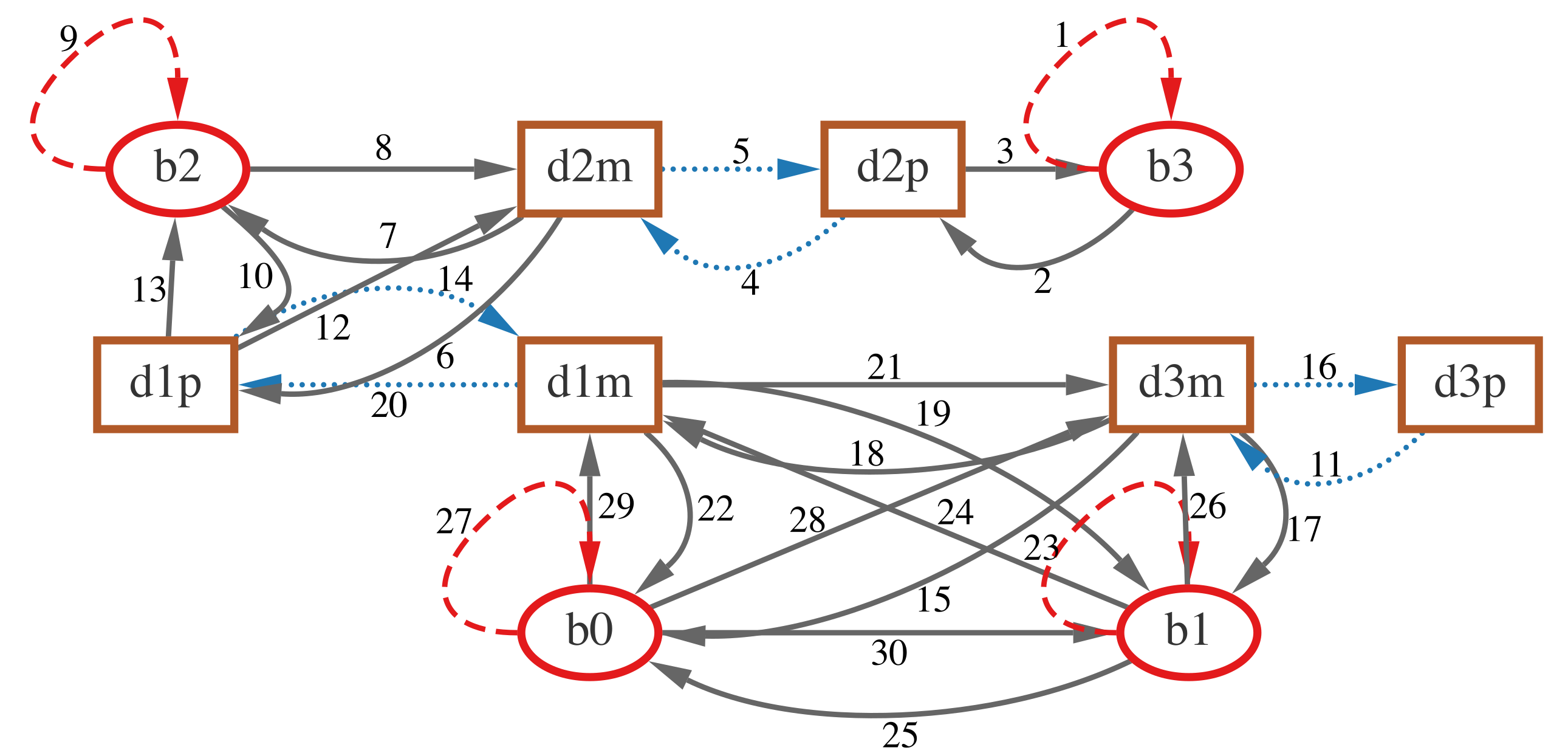}
\caption{\it An EFSM modelling the behavior of the level in Fig. \ref{fig:LR}.
Red circle nodes are in-game buttons; squares are doors. Each door is represented by a pair of nodes, representing the two sides of the door.
The agent starts at $b_0$. 
Solid black transitions represent travel between two game objects. Blue transitions represent travel through a door (only possible if it is open). A red transition represents the toggling of the associated button. The extended state of the EFSM consists of the state of the doors, which is either open or closed (not shown). All doors are initially closed.
Toggling a button will toggle the state of associated doors (not shown). 
}\label{fig:LR.EFSM}   
\end{figure}

Fig. \ref{fig:LR.EFSM}  shows the EFSM that models the level in Fig. \ref{fig:LR}.
Note that the EFSM does not contain information about the physical layout of the world, but only the reachability relation between neighboring entities (doors and buttons). 
With the model at hand, a test case is generated as a sequence of transitions from the initial state. This step happens offline (i.e., without executing on the GUT), greatly speeding up the generation time (minutes, rather than hours). However, the produced test suites cannot be directly executed on the GUT (\lr), since the information on how to navigate through the game world is missing. For instance, a test case could require going from $b_0$ to $b_1$, but the model does not have information on how to navigate in \lr. The \ivxr\ framework fills this gap
by providing the notion of goal structure. In particular, each transition of a test case generated from the EFSM is translated into a goal along with the tactic to solve it. The whole test case, which is a sequence of transitions, is translated to corresponding $\mathsf{SEQ}$ goal structure similar to the example in Section \ref{sec.goalstructure} that can be executed on \lr.

\begin{table}[tb]
\centering
{\scriptsize
\begin{tabular}{lrrr}
    \hline
  & States & Transitions & Variables \\
  \hline
 L1 & 144 & 558 & 40 \\ 
 L2 & 155 & 646 & 40 \\
 L3 & 225 & 1439 & 40 \\
 \hline
\end{tabular}
    \caption{\it Characteristics of the EFSM models corresponding to three randomly generated \lr\ levels. }
    \label{tab:efsm_properties}
    }
\end{table}

\subsection*{Experience}

\begin{figure}[htb]
\includegraphics[scale=1]{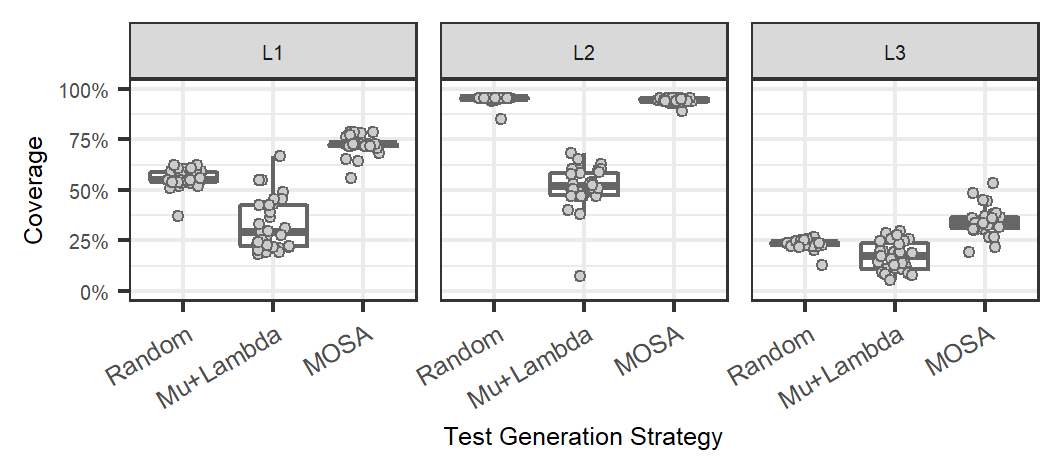}
\caption{\it EFSM transition coverage achieved by different search-based test generation strategies. For each model (L1, L2, and L3), the plot report the boxplot of the transition coverage as well as the coverage (gray dots) of each replica.}\label{fig:LR.COV}   
\end{figure}

An empirical study of this setup has been presented in \cite{ferdous2021search}. Here, we will show a small example to discuss the experience. We automatically generate three large levels, L1, L2, and L3, and the corresponding EFSM models; Table~\ref{tab:efsm_properties} shows their main features. For each model, we take advantage of the search-based test generator tool EvoMBT\footnote{\url{https://github.com/iv4xr-project/iv4xr-mbt}} to assess the performances of three  test suite generation strategies: pure Random generation, classic evolutionary strategy algorithm Mu+Lambda, and many objective algorithm MOSA. For each model, each generation strategy runs 30 times for 300s. EFSM transition coverage observed is reported in Figure~\ref{fig:LR.COV}. 

The EvoMBT tool supports execution on \lr\ of a test suite generated on an EFSM model of a level. EvoMBT exploits the \ivxr\ framework to translate an EFSM transition to a goal structure and to execute it on \lr. For each level and each generation strategy, we selected a test suite generated from the EFSM model and executed it on \lr. As can be seen from Table \ref{tab:mbt_on_sut}, the mean execution time of a test suite on \lr\ is more than two hours and a half, while the generation on the model required only five minutes. Clearly, if test generation were done directly on \lr\, the time required would not be feasible. Another interesting point is the high number of failed tests when run on \lr (mean value 50\%). The execution of a test case fails because the \ivxr\ agent does not complete a goal within a specified time budget or because the agent cannot reach a specific position in the current \lr\ level. The first problem can be solved by increasing the agent's time budget, thus making the execution more time consuming. The second type of failure highlights a problem in the \lr\ autonomous navigation support that we are investigating. 
\begin{table}[h]
\centering
{\scriptsize
\begin{tabular}{llrrrll}
  \hline
  Level & Strategy & Time(h) & n Tests & n Fails \\ 
   &  &  &  &  \\ 
  \hline
 L1 & Random & 2.79 & 108 &   6 \\ 
 L1 & Mu+Lambda & 1.31 &  60 &  52  \\ 
 L1 & MOSA & 2.62 & 160 & 118  \\ 
 \hline
 L2 & Random & 3.54 & 252 & 179  \\ 
 L2 & Mu+Lambda & 0.46 &  33 &  33  \\ 
 L2 & MOSA & 4.15 & 302 & 219  \\ 
 \hline
 L3 & Random & 1.04 &  67 &   9  \\ 
 L3 & Mu+Lambda & 6.05 & 133 &   9  \\ 
 L3 & MOSA & 1.87 &  95 &  23  \\ 
   \hline
\end{tabular}
\caption{\it Exectution of the test suites generated from the EFSM models on \lr. 
\label{tab:mbt_on_sut}
}}
\end{table}

The integration of the \ivxr\ framework with MBT enables fast test suite generation while providing a natural and general notion of coverage based on EFSM models. Here, we presented our experience with \lr\, a real 3D maze game, and we showed that the combination of model base generation and the framework \ivxr\ is effective in identifying potential issues in the game under test. Future work includes supporting other game entities (e.g., fire hazards) in the models, as well as the extension to multi-agent scenarios.

\section{Related Work} \label{sec.relatedwork}

Until today the game industry heavily relies on manual play testing to test games. Automated testing is rarely done. 
The challenges range from process related, where testing is not rigorously implanted in  development cycles, to engineering, e.g. the lack of testing tools that can target games out of the box. 
Politowski et al. give a good overview on issues and challenges of game testing in the industry \cite{politowski2021survey}.

In terms of research, there are indeed work in automated play testing, though in much less volume than in other areas of automated testing.
For example the use of model based testing (MBT) was investigated by Iftikhar et al. \cite{iftikhar2015automated} and later by Ferdous et al. \cite{ferdous2021search}.
However, recent work in automated play testing seem to focus more on the use of machine learning, in particular reinforcement learning (RL). 
E.g. Pfau et al. use RL to train an automated test agent for an adventure game \cite{pfau2017automated}.
Zheng et al. use evolutionary deep RL to do the same for an action game \cite{zheng2019wuji} (which has much more dynamics than an adventure game). 
Ariyurek et al. use RL to train an agent to play with different styles, e.g. killer or explorer \cite{ariyurek2019automated}. Gordillo et al. use curiosity driven RL to improve coverage.

Despite advances in RL, its scalabilty for game testing is still an open discussion. RL requires a huge amount of training, which all must be executed on the actual game where actions are relatively much slower to execute. So, the overall computation cost might be prohibitive e.g. for smaller companies. Moreover, if the game logic is changed, or the world layout is changed, which happen very often during the development, the agent may have to be retrained.

Obviously programming a game play is much harder than, for example, scripting a test sequence for a web application. For this reason the fascination towards RL is understandable. However we can also reduce the investment cost for building automation by providing a proper programming language, or at least a framework, that offers the right concepts and abstraction so that the effort for programming play testing becomes manageable. The \ivxr{} framework tries to fill this role. The advantage of a more programming-based approach is that we have much more control on the test agent behavior, and we have also shown that a BDI test agent is robust against development time changes \cite{shirzadehhajimahmood2021}.

\section{Conclusion} \label{sec.concl}

We have discussed our experience of using \ivxr{} to do automated testing on three different games, ranging from a turn-based 2D game to a complex commercial 3D game. In all three cases the use of \ivxr{} has successfully introduced automation and contributed in finding bugs and issues. Unlike e.g. a machine learning based approach, \ivxr{} is a programming approach that allows automated testing to be programmed at a high level. The approach gives developers more control on the behavior of the test agent while keeping the agent versatile and robust.
Moreover, we have demonstrated that \ivxr{} can be used as a rich interface to enable more traditional testing tools to target computer games.
Building an interface between the game under test and \ivxr{} along with a library of basic tactics and goals
does require some effort, but this is one off investment, after which developers will benefit from powerful test automation.

\begin{acks}
This work is supported by the \grantsponsor{H2020SponsorIDblabla}{EU ICT-2018-3 H2020 Programme}, grant nr. \grantnum{H2020SponsorIDblabla}{856716}.
\end{acks}


\bibliographystyle{ACM-Reference-Format}
\bibliography{mybib}
\end{document}